 \documentclass[final,3p,times,12pt]{elsarticle}

\usepackage{amssymb}

\usepackage{subfigure}          

 \usepackage{lineno}

 \makeatletter
\def\ps@pprintTitle{%
  \let\@oddhead\@empty
  \let\@evenhead\@empty
  \def\@oddfoot{\reset@font\hfil\thepage\hfil}
  \let\@evenfoot\@oddfoot
}

\journal{Astroparticle Physics}

\begin{document}


\begin{frontmatter}



\title{The muonic longitudinal shower profiles at production}

 \author[LIP]{S. Andringa}
 \author[LIP]{L. Cazon}
 \author[LIP]{R. Concei\c{c}\~{a}o\corref{cor1}}
 \ead{ruben@lip.pt}
 \cortext[cor1]{Corresponding author}
 \author[LIP,IST]{M. Pimenta}
 \address[LIP]{LIP, Av. Elias Garcia, 14-1, 1000-149 Lisboa, Portugal}
 \address[IST]{Departamento de F\'{i}sica, IST, Av. Rovisco Pais, 1049-001 Lisboa, Portugal}

\begin{abstract}

In this paper the longitudinal profile of muon production along the shower axis is studied. The characteristics of this distribution is investigated for different primary masses, zenith angles, primary energies, and different high energy hadronic interaction models. It is found that the shape of this distribution displays universal features similarly to what is known for the electromagnetic profile. The relation between the muon production distribution and the longitudinal electromagnetic evolution is also discussed.

\end{abstract}

\begin{keyword}
Extensive Air Shower \sep Longitudinal Profile \sep Muon Production Depth \sep Electromagnetic component

\end{keyword}

\end{frontmatter}


\section{Introduction}
\label{sec:Intro}

The origin and mass composition of the most energetic particles in the Universe, the Ultra High Energy Cosmic Rays (UHECRs), remains a mystery. These particles reach the Earth with a very scarce flux. Fortunately, their interaction with the atmosphere molecules produces huge cascades of particles, known in the literature as Extensive Air Showers (EAS). The detection of these showers is usually done by measuring the charged particles that arrive at the ground, or, in moonless nights, the development of the shower can be followed through the fluorescence light produced by the EAS.

While the arrival direction of the UHECRs can be easily obtained using any of the techniques mentioned, the mass composition of the primary particle is much more difficult. The shower observables connected to the type of primary particle are also sensitive to the physical interactions that occur during shower development. The hadronic interactions at high energies are described through phenomenological models that are fitted to the available accelerator data and extrapolated several orders of magnitude to the UHECRs energies. Moreover, the accelerators have difficulty to reach the most important region for the characterization of the EAS development, the forward region, increasing the uncertainties on the extrapolation.

Contrary to the fluorescence light, which is dominantly produced by low energy electrons in secondary electromagnetic cascades, ground signals are sensitive to muons produced at different depths, thus imaging the hadronic cascade. 

Many of the muons decay before reaching the ground. Current EAS Monte Carlo simulations (for instance CORSIKA \cite{CORSIKA} or AIRES \cite{AIRES1}) properly account for the propagation effects starting from the moment of production. The \emph{total/true} Muon Production Depth distribution (MPD) is defined as the total number of muons produced in each slant depth unit, regardless of the probability to reach ground and be detected. On the other hand, the \emph{apparent}-MPD distribution is affected by the geometrical and propagation effects and it has been extensively studied \cite{referee1_1,referee1_2,referee2_1,referee2_2,referee2_3,MPDAuger}.
In \cite{TransportModel} the \emph{apparent} is related to the \emph{true} MPD distribution through the energy, transverse momentum at production and propagation effects. This knowlegde would allow the reconstruction or at least constrain the \emph{total/true} MPD, which is closer to the development of the hadronic cascade, i.e. without being masked by propagation effects.

The absolute density of muons at ground has been used previously to study high energy hadronic interaction models and primary composition of cosmic rays. The number of muons at a given distance from the core is analised in \cite{muonAuger}, showing that at very high energy the data can not be explained by the available models.
The simultaneous analysis of the electromagnetic calorimetric energy and ground signals has been used in \cite{invEAuger} to correct for the \emph{invisible} energy carried by neutrinos (and muons), showing that the energy determination can be improved on an event-by-event basis.

In this paper we identify the main characteristics of the \emph{total/true} MPD distribution characteristics at UHECRs energies. 
The shape of the depth profile, depth of maximum, $X^\mu_{max}$, and its corresponding value, $N^\mu_{max}$ will be studied. The comparison with the features observed in the energy deposit profile, referred to in this paper as electromagnetic profile, will be made whenever relevant.

Since in this study we are only interested on the longitudinal profile, and not the transverse distributions, we used the hybrid shower simulation CONEX \cite{conex1,conex2} (version v2r3). This program combines Monte Carlo simulation with one-dimensional cascade equations making it very fast and allowing the production of large samples of showers. CONEX gives as output the energy deposit profile as a function of depth ($X$), the number of muons produced as a function of $X$, and the more extensively used number of muons along the shower axis. Note that this last one, the number of muons as a function of the shower depth, is in first approximation the cumulative of the number of muons produced, minus muons that decay. Instead we will concentrate our study in the muon production profile.

In this work we study the dependence of the muon shower production profile on: the primary mass composition (proton and iron) at several energies; the zenith angle at $E=10^{19}$ eV; different hadronic interaction models, in particular QGSJet-II.03 \cite{QGSII1,QGSII2} and EPOS1.99 \cite{EPOS} at fixed energy $E=10^{19}$; the primary energy (from $\log(E/{\rm eV}) = 17.5\ - \ 20.0$ in steps of $0.5$). For each different set of parameters samples of $50\,000$ showers were generated. The high energy hadronic interaction model used as default was QGSJet-II.03. The ground was set at $4000$ ${\rm g\,cm^{-2}}$ to avoid an abrupt termination allowing us to see the full tail of the shower profile as if it had a very inclined zenith angle, $\theta$.

The paper is organized as follows: in Sec. \ref{sec:long} the main features of the muon production shower profile are studied and related to the electromagnetic profile; the shape of the profile is characterized in Sec. \ref{sec:universality} and here it is shown that, similarly to what happens in the electromagnetic case, this profile exhibits an universal behavior when expressed in the coordinates, $X' \equiv X - X_{max}$ and $N' \equiv N/N_{max}$; in section \ref{sec:Ecut} the impact of the muon energy threshold on the longitudinal profile is investigated; the extraction of more information on basic variables like the total number of muons or the point of first interaction are discussed in Sec. \ref{sec:NbMuons}; the paper ends with conclusions and prospects.

\section{Longitudinal shower profiles}
\label{sec:long}

\begin{figure}[htbp]
\begin{center}
\includegraphics[width=0.8\textwidth]{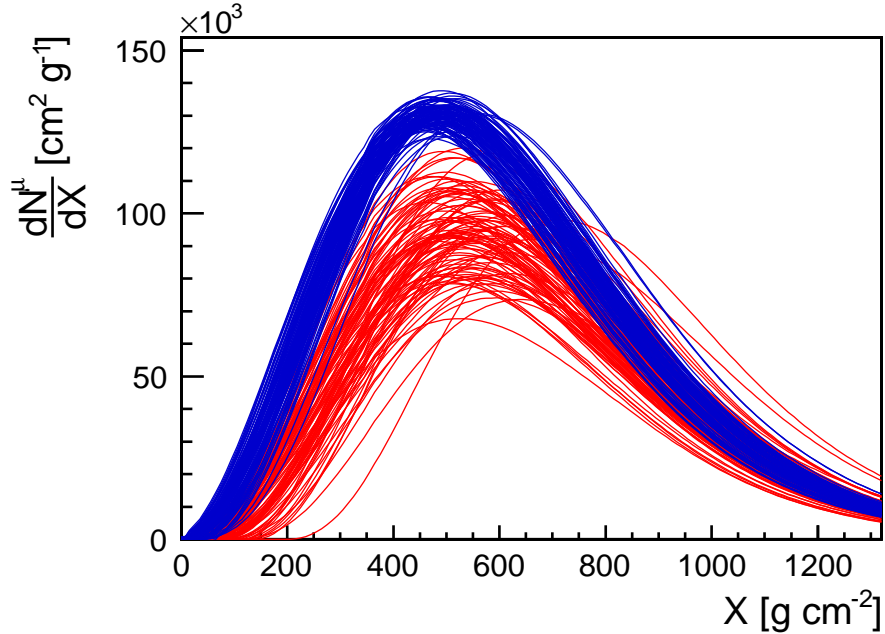}
\caption{Muon production shower profiles as a function of depth ($X$), for proton (red) and iron (blue) primaries at $E = 10^{19}$ eV. In this picture is shown $100$ showers for each primary, both generated with QGSJet-II.03 and with $\theta = 40^\circ$.}
\label{fig:prof}
\end{center}
\end{figure}

\begin{figure}[htbp]
 \begin{center}
  \subfigure[] {
   \includegraphics[width=0.45\textwidth]{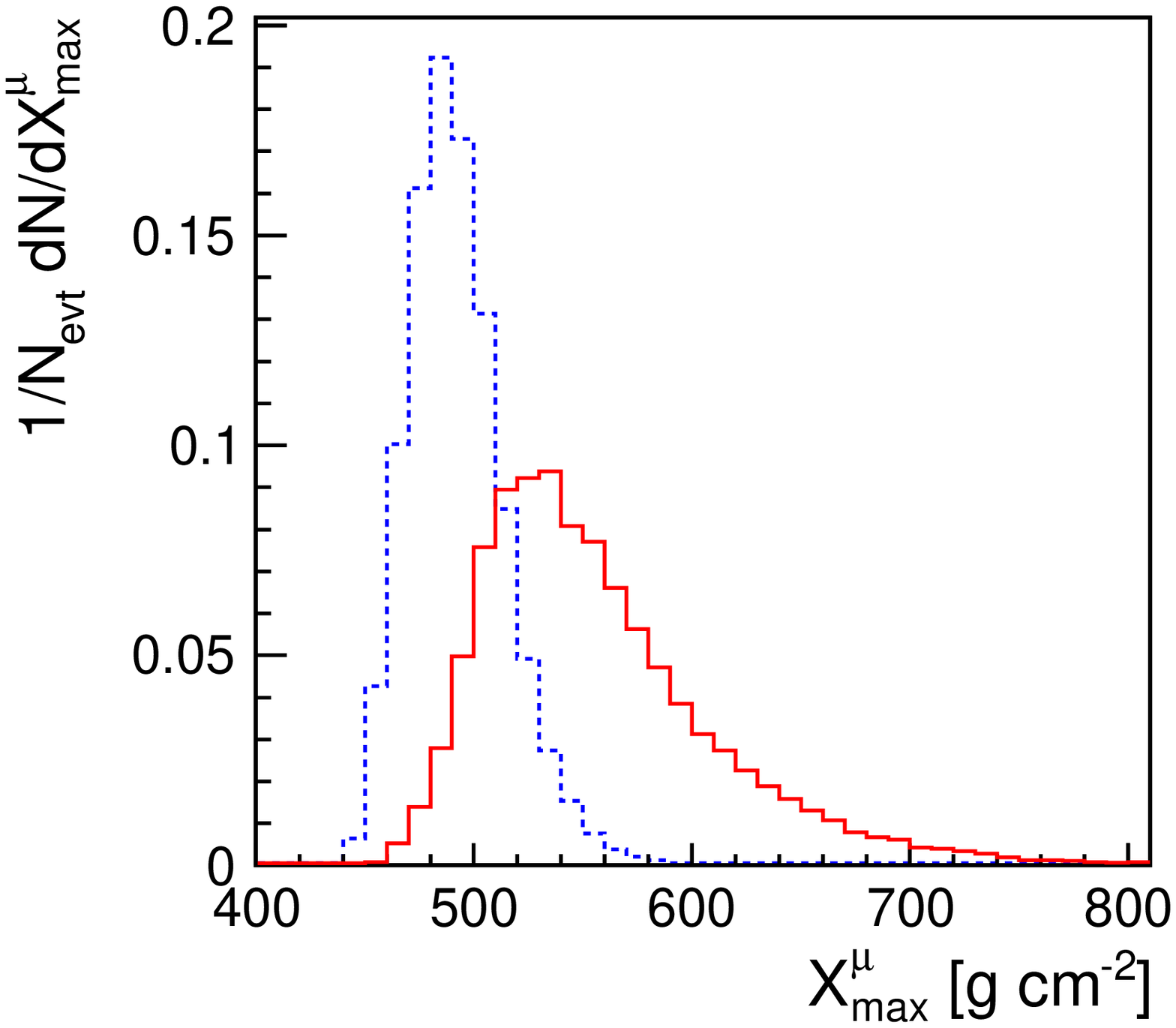}
    \label{fig:XmaxmuC}
  }
   \hspace{0.7cm}
  \subfigure[] {
   \includegraphics[width=0.45\textwidth]{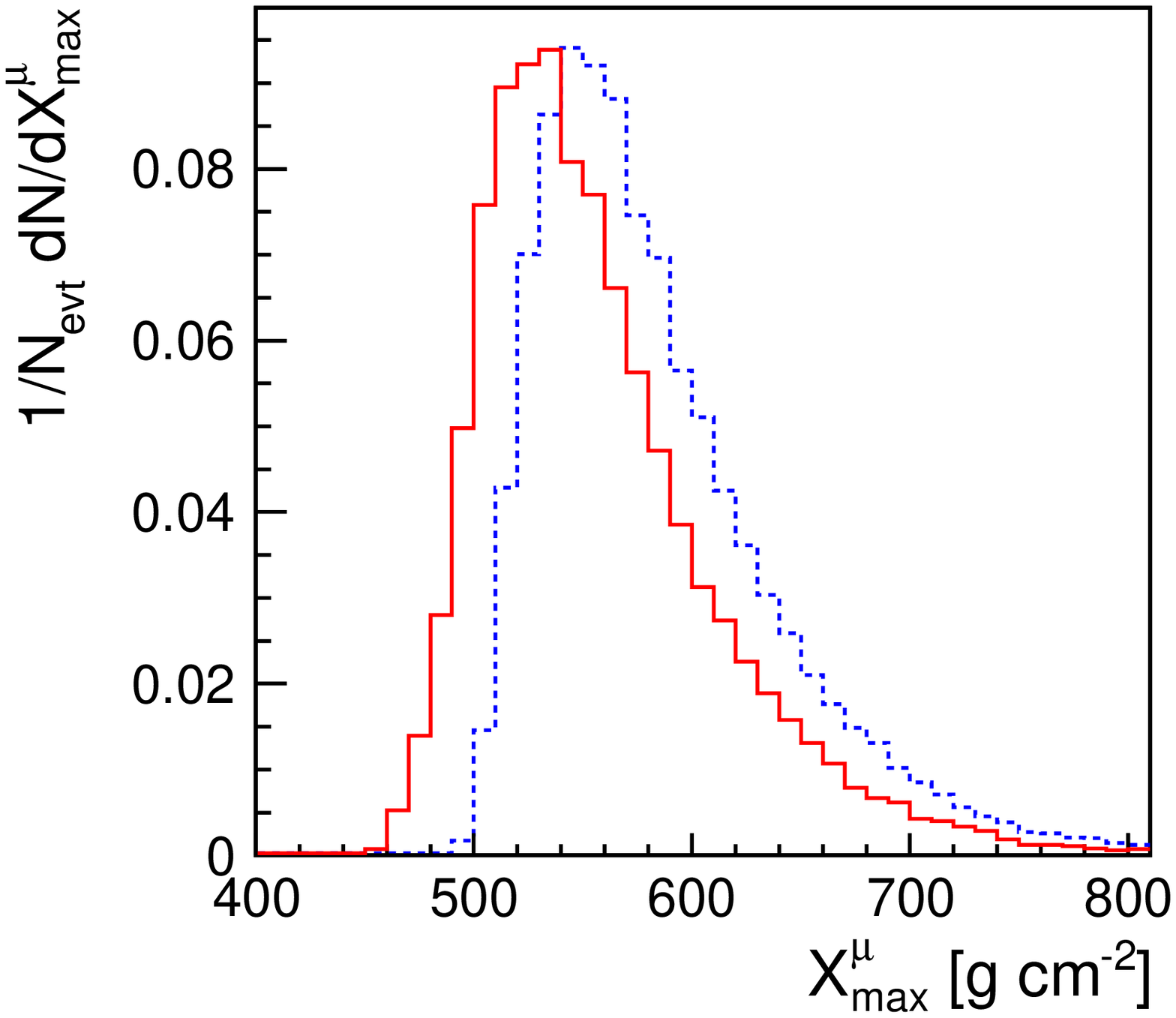}
    \label{fig:XmaxmuH}
  }
   \caption{$X^{\mu}_{max}$ distributions of muon production longitudinal profiles for: (a) different primaries - the proton distribution is in red/full while iron is in blue/dashed (showers generated with QGSJet-II); (b) high energy hadronic interaction models - QGSJet-II is the red/full line and EPOS1.99 is shown as blue/dashed (for proton induced showers).}
   \label{fig:Xmax}
 \end{center}
\end{figure}

The muon production longitudinal profile for proton induced showers (in red) and iron primaries (blue) at $E = 10^{19}$ eV is shown in figure \ref{fig:prof}. These profiles were obtained with CONEX where the minimum energy threshold for muons is $1$ GeV (the effect of this cut is discussed in Sec. \ref{sec:Ecut}). 

The shower profiles reflect the properties of the first stages of the hadronic interaction, in particular the first one, which can not be observed directly. They are thus very different for proton and iron initiated showers, as shown in Fig. \ref{fig:prof}. 
Compared to proton showers, iron showers have a higher number of muons, which is readily seen just by looking at the maximum of the profile\footnote{in this paper the index $\mu$ and $e.m.$ will be used to address to variables related with the muon production profile and the electromagnetic profile, respectively.}, $N^{\mu}_{max}$.In iron showers more charged pions are produced and as a consequence it has more muons. Low energy pions will give rise to muons as they decay, while the pions above the critical energy\footnote{the critical energy is defined as the energy where the decay lenght equals the interaction length.} are more likely to interact producing more pions.
Protons exhibit much more fluctuations: this is due to the smaller pion multiplicity in the first interactions, thus less muons are produced and these are more spread along the shower axis.

\begin{figure}[htbp]
 \begin{center}
  \subfigure[]{
   \includegraphics[width=0.45\textwidth]{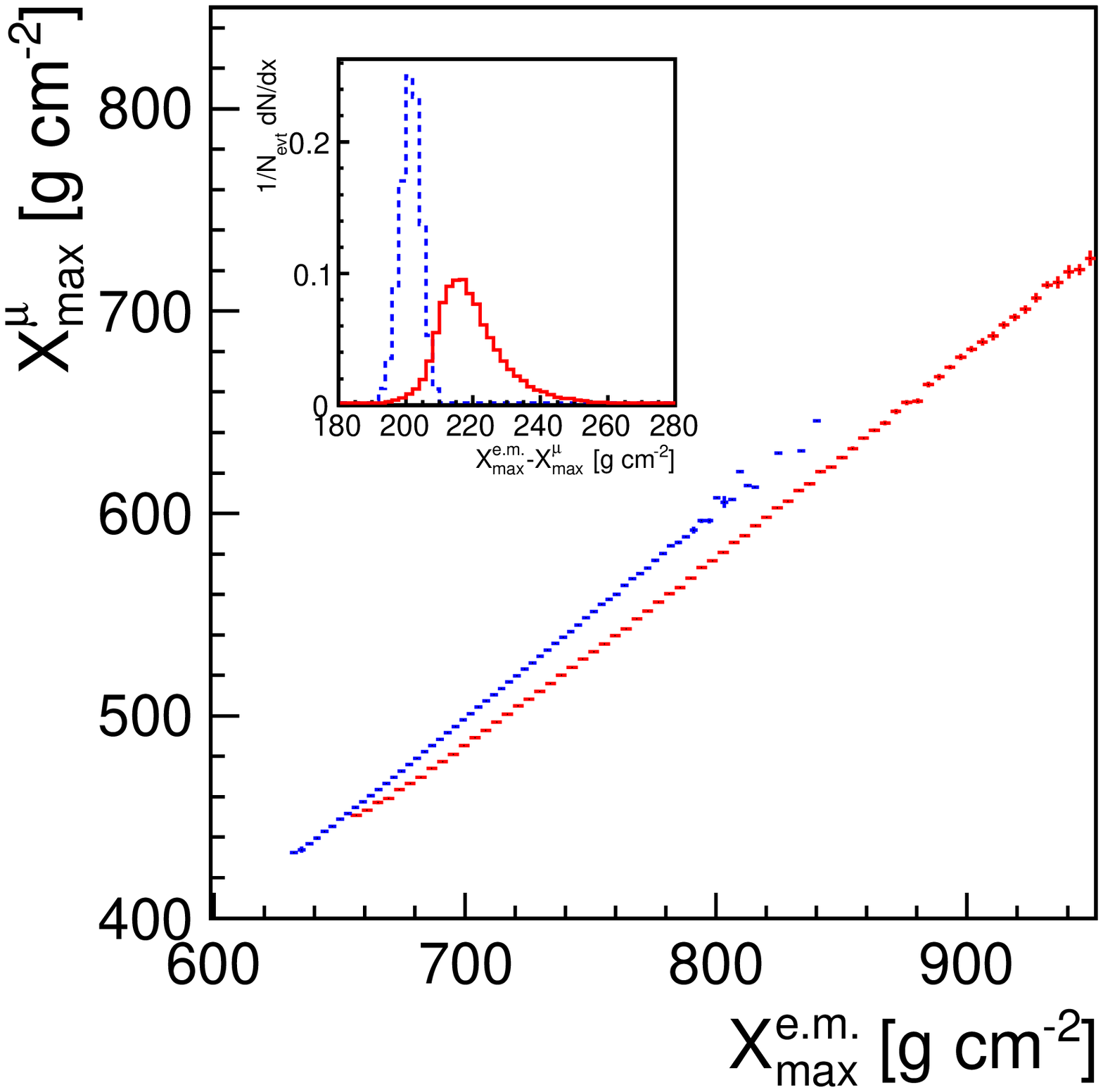}
    \label{fig:Xmaxem-muC_prof}
  }
   \hspace{0.7cm}
  \subfigure[]{
   \includegraphics[width=0.45\textwidth]{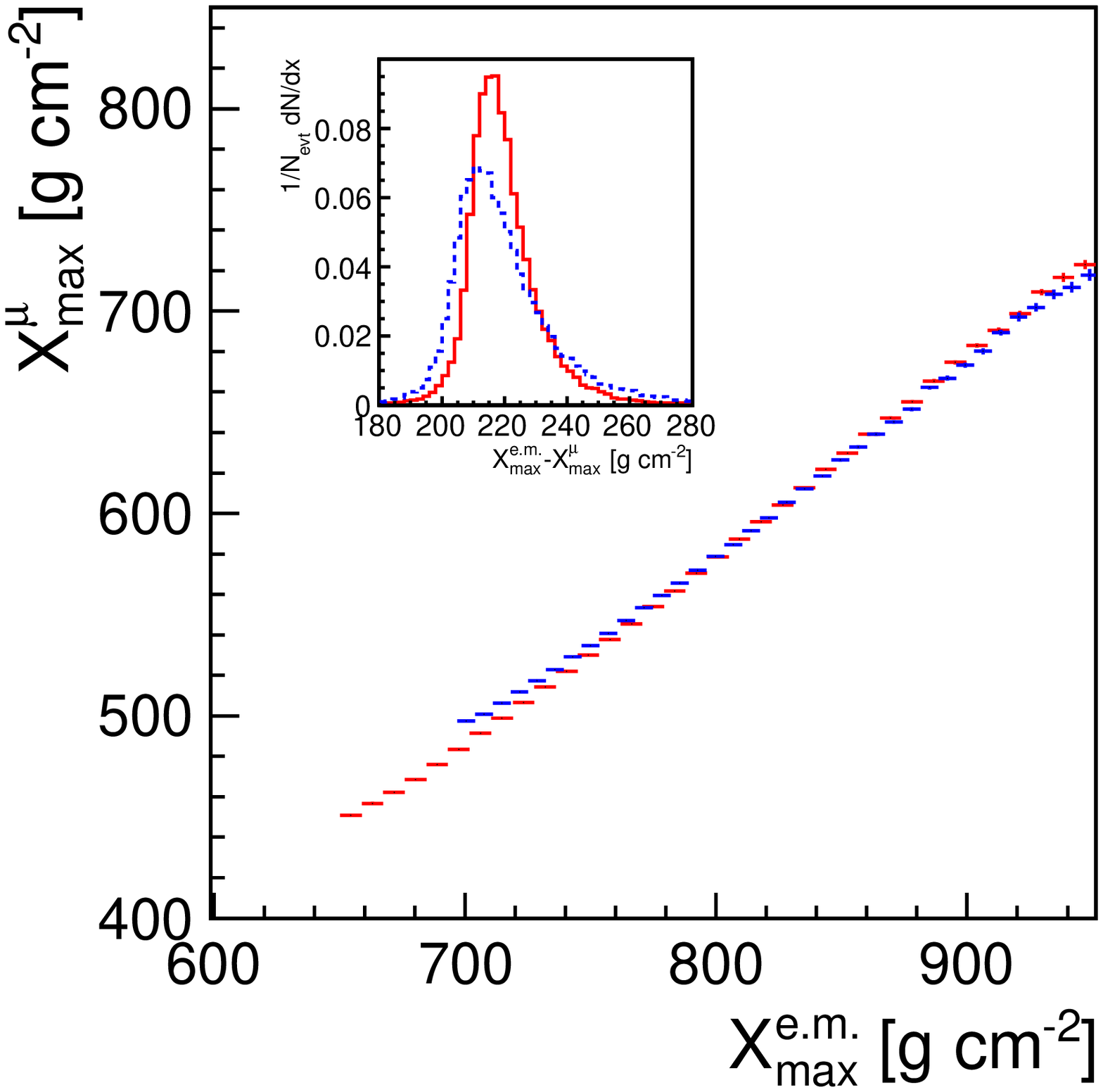}
    \label{fig:Xmaxem-muH_prof}
  }
   \caption{Relation between muon production $X^{\mu}_{max}$ and electromagnetic $X^{e.m.}_{max}$, for $E = 10^{19}$ eV showers. The difference between the two $X_{max}$ are shown in the inset plots. In (a) the dependence on primary mass is shown (proton is the red (full) line and iron the blue (dashed) line, both generated with QGSJet-II), while in (b) the dependence on high energy hadronic interaction models for proton induced showers (QGSJet-II is the red (full) line and EPOS1.99 the blue (dashed) line).}
   \label{fig:Xmaxmu}
 \end{center}
\end{figure}

Neutral pions of all energies decay immediately and around 90\% of the primary energy will feed electromagnetic cascades; the energy deposit in the atmosphere is dominated by low energy electrons after a few radiation lengths. 
All of the sub-cascades add up to give a well defined maximum height of the profile. Not only the integral of the profile is a good calorimetric measurement of the energy of the primary particle (which must be corrected for the muons and accompanying neutrinos which reach the ground), but also the maximum of the profile is directly proportional to the energy, within 5\% \cite{USPV} for both primaries. Because it corresponds to a much higher number of particles than those in the muon production profile this number has much less fluctuations. 
In the electromagnetic case, it is the depth of the maximum, $X^{e.m.}_{max}$, that gives more information about the other shower properties.

The depth of the maximum of a shower, $X_{max}$, is determined by the depth of first interaction, $X_1$, but also the subsequent development, $\Delta X$. 
The variation of the first is common to both the electromagnetic and muon profiles, while the later is different. 
Iron primaries have a larger cross-section and higher multiplicity and both effects contribute to make the average maximum depth, $\left< X_{max} \right>$, and $RMS(X_{max})$ smaller than the corresponding proton values.
In Fig. \ref{fig:Xmax}, we show the $X^{\mu}_{max}$ distributions for the muon production profiles, comparing different primaries and different hadronic interaction models. 
The features are similar to the ones of the electromagnetic $X^{e.m.}_{max}$, and the correlations between muonic and electromagnetic maxima can be seen in Fig. \ref{fig:Xmaxmu}, together with the differences.

The electromagnetic maximum is reached around $200$ ${\rm g\,cm^{-2}}$ later, after the energy is degraded to a large number of electrons (which will then stop multiplying), while the muonic profile has a maximum when the hadronic shower is still important. 
There is a correlation between the electromagnetic and the muonic $X_{max}$, shown in fig. \ref{fig:Xmaxmu}. It depends slightly on the primary mass and is almost independent of the hadronic interaction models.

The difference between electromagnetic and muonic $X_{max}$, shown as an inset plot of fig. \ref{fig:Xmaxmu}, also changes between primaries -- by around 30 ${\rm g\,cm^{-2}}$ in the average separation -- which means that although related, the electromagnetic and muonic profiles give some independent information.
Notice that this difference, $X^{e.m.}_{max} - X^{\mu}_{max}$, is only sensitive to $\Delta X$, thus it is giving information about the history of pion production.

\section{Universality of the longitudinal profile shape}
\label{sec:universality}

\begin{figure}[htbp]
 \begin{center}
   \includegraphics[width=0.8\textwidth]{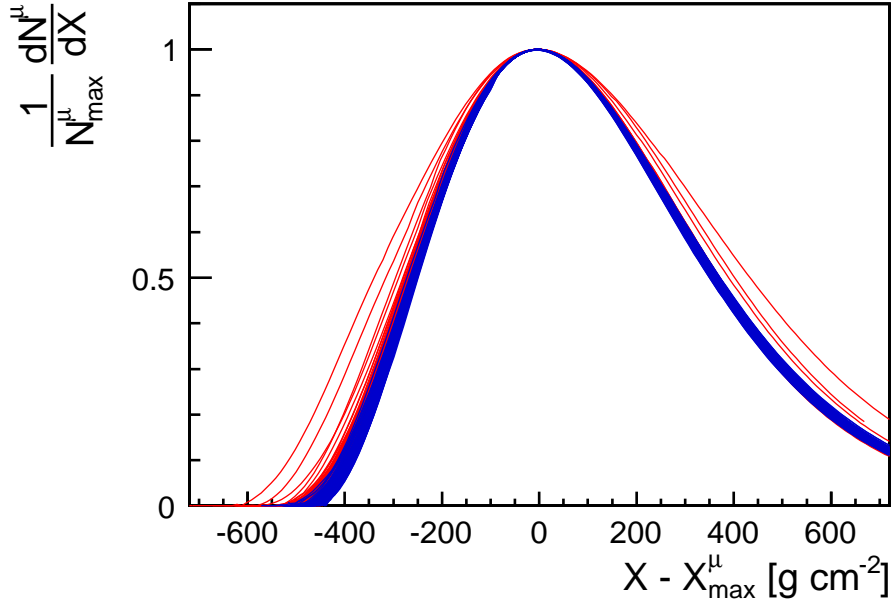}
  \caption{Muon production shower profiles from proton (red) and iron (blue) primaries, in $(X',N')$ coordinates. The same showers used to build Fig. \ref{fig:prof} are used here.}
   \label{fig:uspprof}
 \end{center}
\end{figure}

\begin{figure}[htbp]
 \begin{center}
   \includegraphics[width=1.00\textwidth]{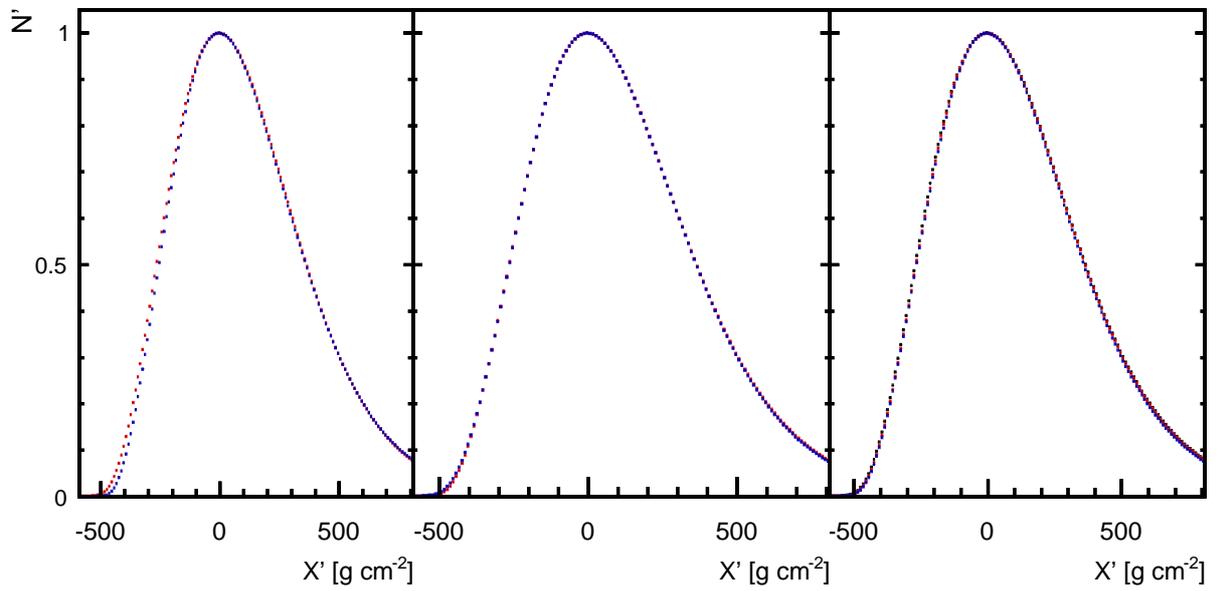}
    \label{fig:uspdmuC}
   \caption{Average muon production longitudinal profile in $(X',N')$ coordinates: (left) dependence on primary mass - proton (red) and iron (blue), showers generated with QGSJET-II at $\theta = 40^\circ$; (middle) dependence on the hadronic interaction model - QGSJet-II (red) and EPOS1.99 (blue), for proton induced showers at $\theta = 40^\circ$; (right) dependence on the zenith angle, $\theta$ - Black for $0^\circ < \theta < 10^\circ$, red for $30^\circ < \theta < 40^\circ$, and blue for $45^\circ < \theta < 55^\circ$, for proton showers using QGSJet-II.}
   \label{fig:uspC}
 \end{center}
\end{figure}

In Fig. \ref{fig:uspprof}, the profiles are expressed in $X' \equiv X - X_{max}$ and $N' \equiv N / N_{max}$. 
The obtained shape is rather universal, similarly to what happens to the energy deposit profile. It can be useful to use the average shape in order to determine the two main parameters from a fit with a reduced set of data. 
On the other hand, there can be extra information on the shape, and we can now construct average profiles to look in detail for differences between primaries and hadronic interaction models. 
From that comparison, in fig. \ref{fig:uspC} (left), it is clear that iron showers develop faster, with almost no difference between models. 

The dependence on zenith angle is also studied. 
Muon production depends on the competition between the pion interaction and decay lengths. 
The larger the energy the most probable it is for a pion to interact instead of decaying into a muon, but that dependence can not be expressed in $X$ alone, as the decay length is independent of the atmospheric density.
Nevertheless, the average muon profile, in figure \ref{fig:uspC} (right), is still rather universal. 
There are, as expected, some minor violations to this universality that are more important in the early phase of the shower.

\begin{figure}[htbp]
 \begin{center}
   \includegraphics[width=0.8\textwidth]{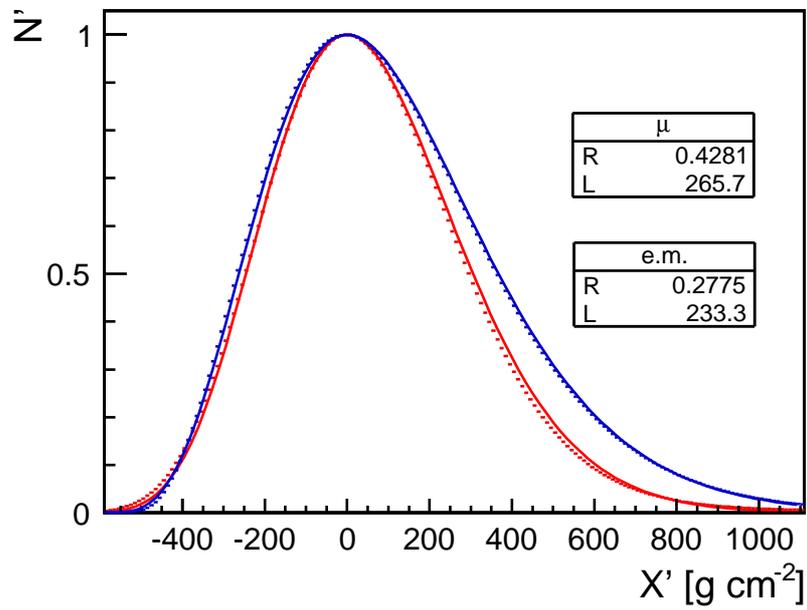}
   \caption{Average shower profiles for proton primaries at E = $10^{19}$ eV, with QGSJet-II, in $(X',N')$ coordinates. Comparison between electromagnetic (in red) and muonic (in blue) shape features. The lines correspond to fits using a Gaisser-Hillas function (2 parameters). The fit results are given in the plot for the electromagnetic (e.m.) and muonic ($\mu$) profiles.}
   \label{fig:profemdmuusp}
 \end{center}
\end{figure}

This quasi-universal shape is compared to the electromagnetic shape in fig. \ref{fig:profemdmuusp}.
The differences are clear: the muonic profile has a steeper growth and is more asymmetric, with respect to the shower maximum.
Both profiles are fitted with a Gaisser-Hillas function, written in terms of a Gaussian width $L$ and a parameter, $R$, which is related with the asymmetry of the shower in respect to the shower maximum \cite{USPV},

\begin{equation}
        N' = \left( 1 + \frac{ R X' }{ L} \right)^{R^{-2}} \exp \left( - \frac{ X' }{ LR } \right).
        \label{USPVform}
\end{equation}

\begin{figure}[htbp]
 \begin{center}
  \subfigure[]{
   \includegraphics[width=0.45\textwidth]{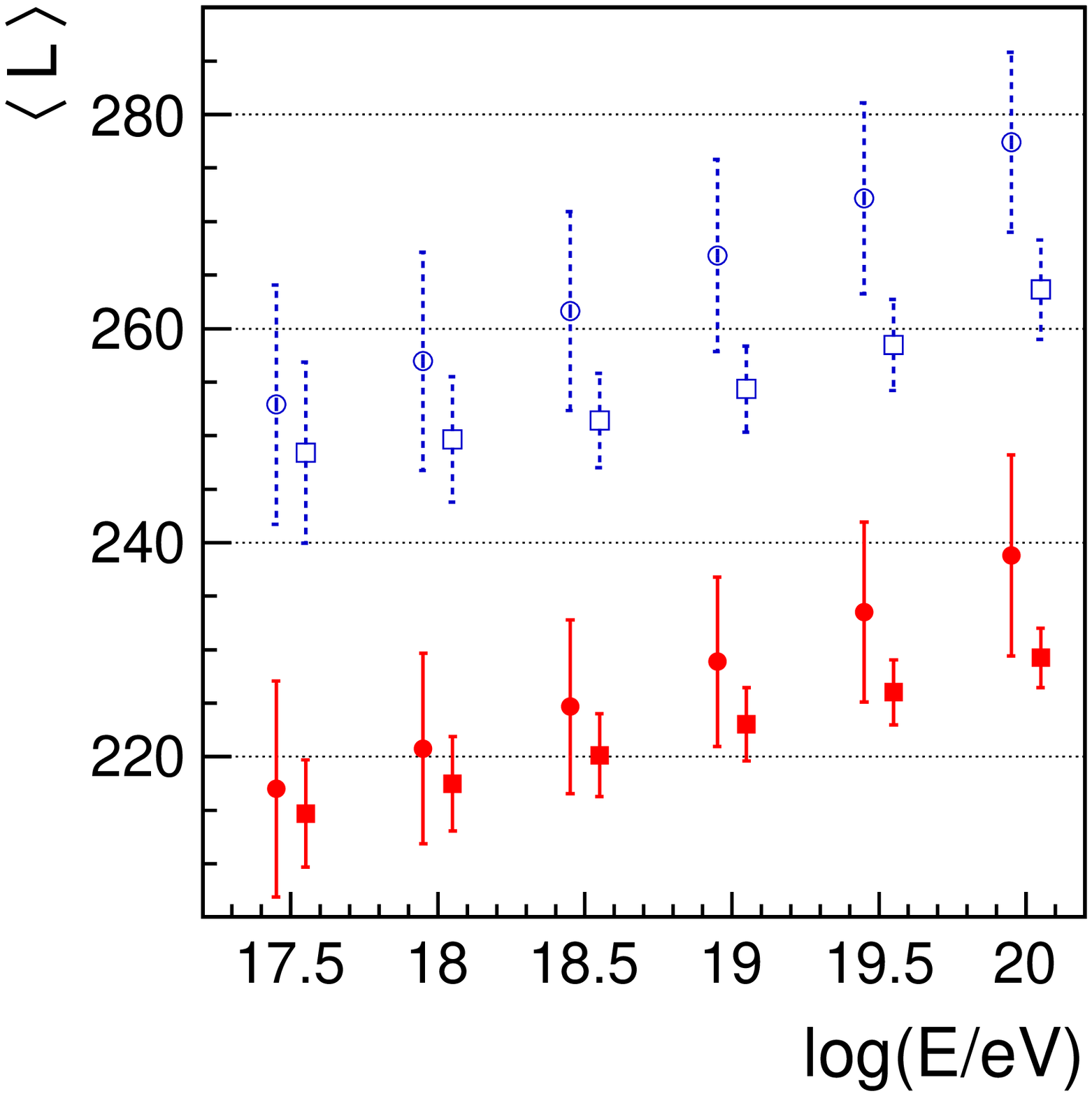}
    \label{fig:uspem}
  }
   \hspace{0.7cm}
  \subfigure[]{
   \includegraphics[width=0.45\textwidth]{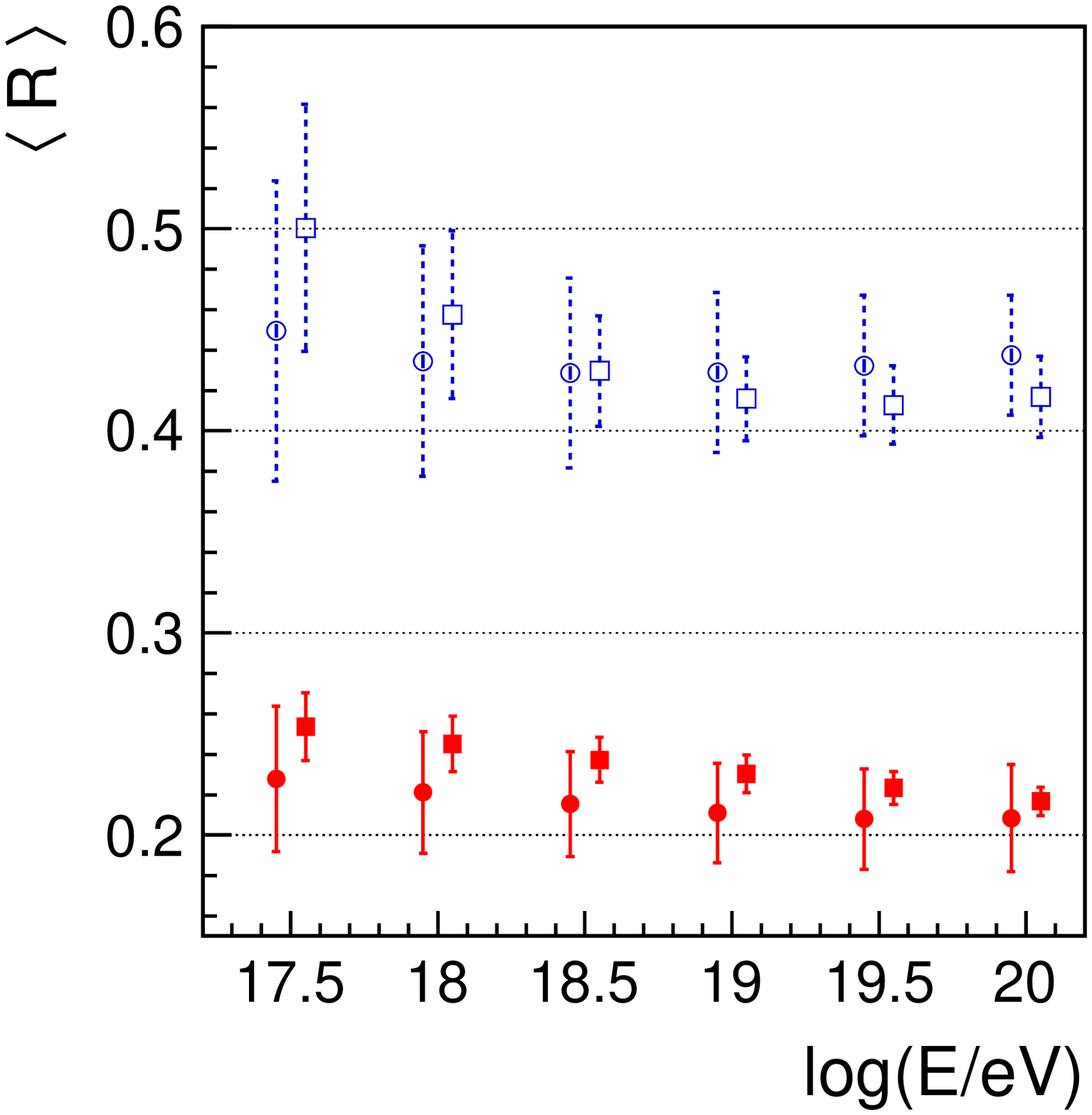}
    \label{fig:uspdmu}
  }
   \caption{Shape parameters dependence on the shower energy. In (a) is shown the $L$ parameter while in (b) is the results for $R$. The shape parameters for the electromagnetic profile are shown in red (full) line while the muon production is shown in blue (dashed). The circles correspond to proton induced showers and the squares have as primary particle iron. The error bars represent the RMS of the corresponding distribution. The points were artificially displaced for better visualization (proton $\log(E/eV) = -0.05$ and iron  $\log(E/eV) = +0.05$. The showers were generated using QGSJet-II as high energy hadronic interaction model and with $\theta = 40^\circ$.}
   \label{fig:usp_evol}
 \end{center}
\end{figure}

We conclude that the muon production profile can be well described with the same function as used for the electromagnetic profile, as seen in figure \ref{fig:profemdmuusp}. It should be noted that the region around the maximum is better described by a Gaisser-Hillas function for the electromagnetic profile. However, the full profile description is better achieved through the Gaisser-Hillas parametrization for the muon production profile. This is because the end tail of the energy deposit profile has important indirect contributions from muon decays. For showers initiated by protons of $10^{19}$ eV, the width of the average muon production profile is larger by $40$ ${\rm g\,cm^{-2}}$ and the asymmetry almost the doubled with respect to the electromagnetic profile. 
The average values and dispersion of these parameters evolve slowly with $\log(E)$, as shown in fig. \ref{fig:usp_evol}, for both primaries (proton and iron). 

\begin{figure}[htbp]
 \begin{center}
  \subfigure[]{
   \includegraphics[width=0.45\textwidth]{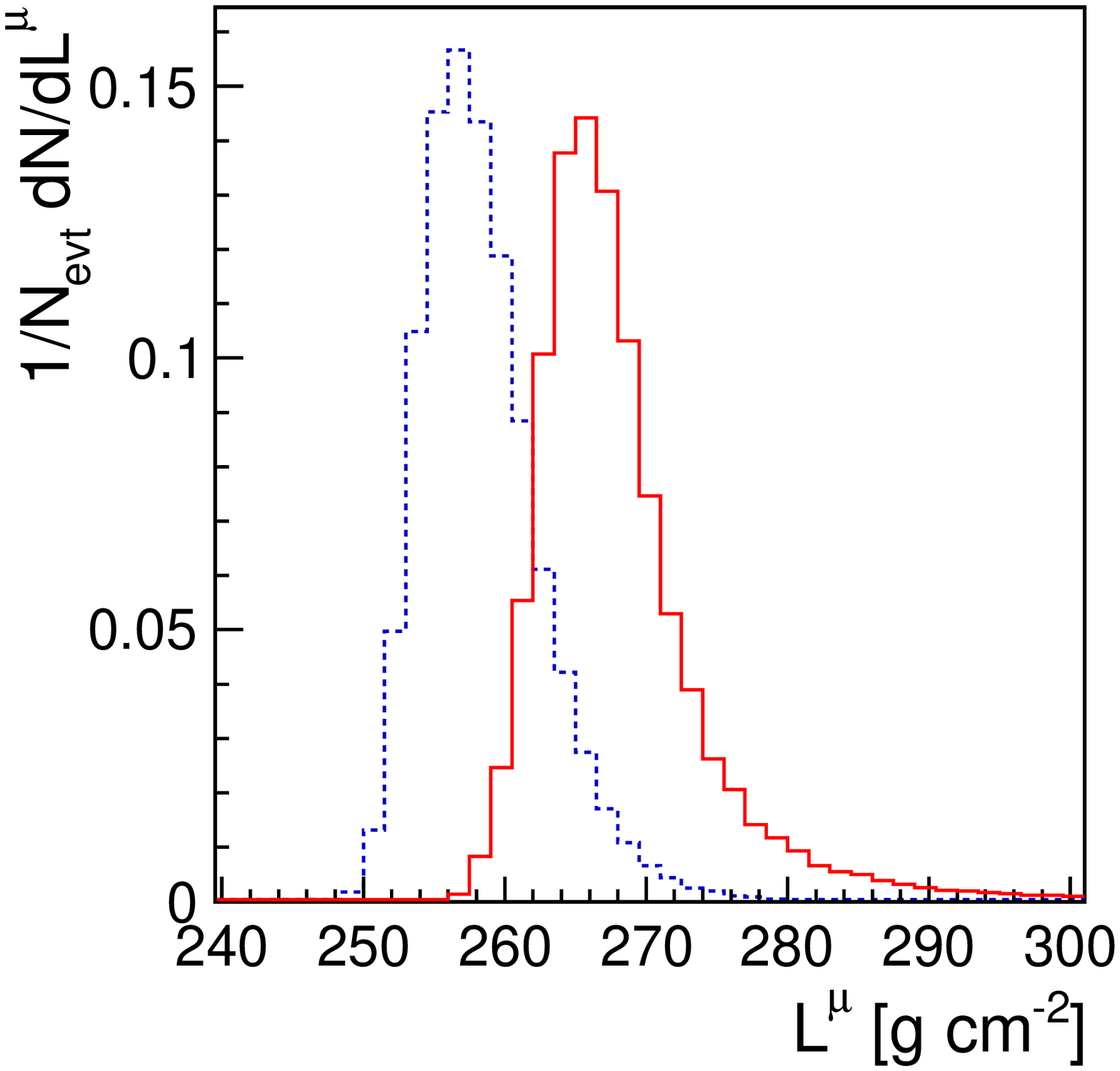}
    \label{fig:uspdmu1}
  }
   \hspace{0.7cm}
  \subfigure[]{
   \includegraphics[width=0.45\textwidth]{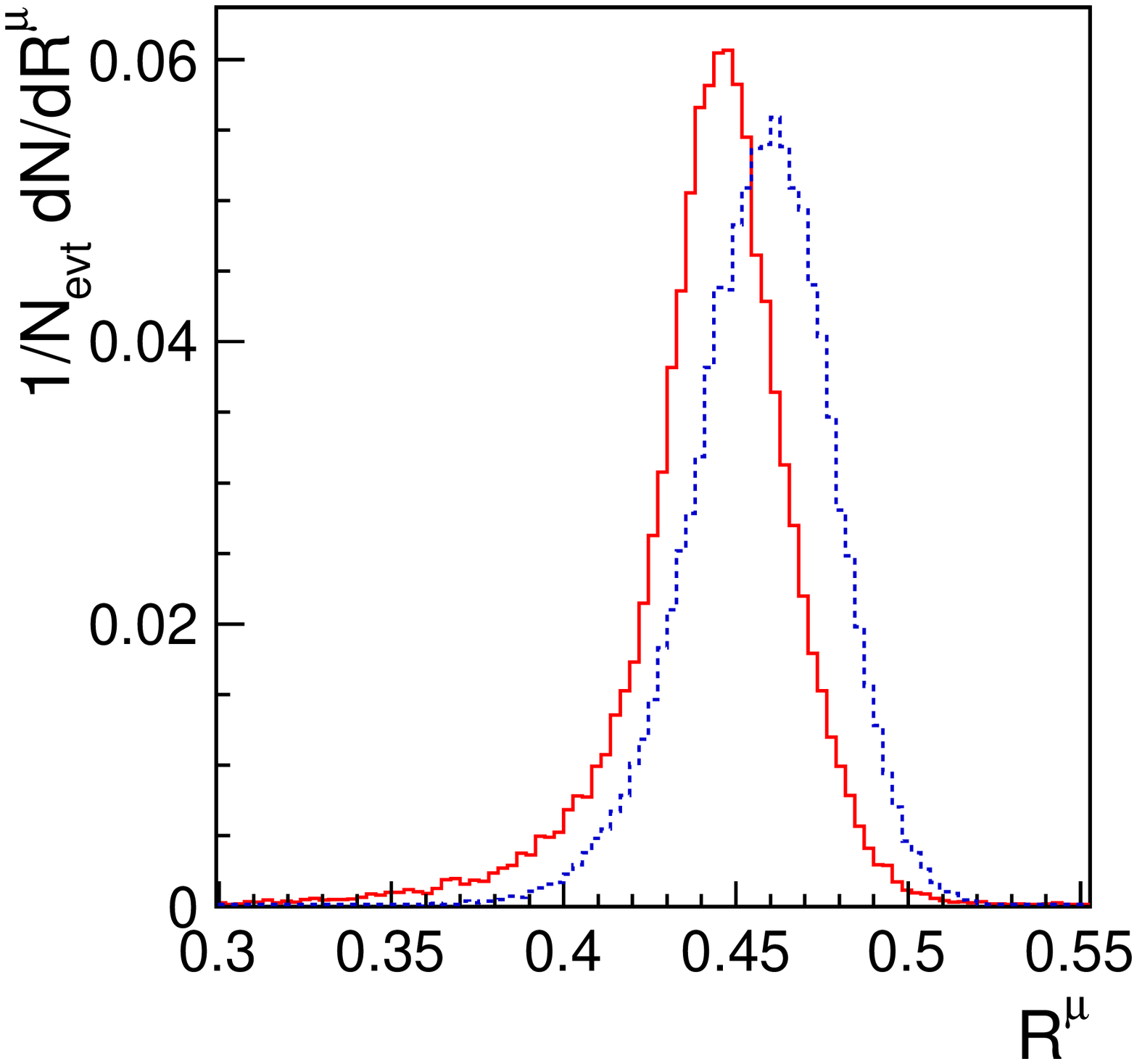}
    \label{fig:uspdmu2}
  }
   \caption{Shape parameters of the muon production profile for different primaries: proton (red/full) and iron (blue/dashed), at E = $10^{19}$ eV, and with QGSJet-II. In (a) is shown $L^{\mu}$ parameter distribution while $R^{\mu}$ is shown in (b).}
   \label{fig:uspp}
 \end{center}
\end{figure}

The differences found for proton and iron profile shapes can now be quantified in these parameters, with the corresponding distributions shown in fig. \ref{fig:uspp}. 
The asymmetry is almost the same for both primaries, the difference in means being well below the 5\% dispersion in each sample. 
Since $R^{\mu}$ only affects the tails it can easily be fixed in the following analysis.
The Gaussian width, on the other hand, has different means for each primary, well above the single primary dispersion, and consistently for the two hadronic interaction models studied. So, it is a new variable for mass composition studies, and fairly model independent.


Clearly $L^{\mu}$ is giving information about $\Delta X^{\mu}$, as it has been obtained starting from $X-X^{\mu}_{max}$, with no memory of $X_1$. It can be measured with ground detectors only.
$L^{\mu}$ can be combined with $X^{\mu}_{max}$ to obtain $X_1$ similarly to what is made for the electromagnetic case \cite{USPV}.




\begin{figure}[htbp]
 \begin{center}
   \includegraphics[width=0.8\textwidth]{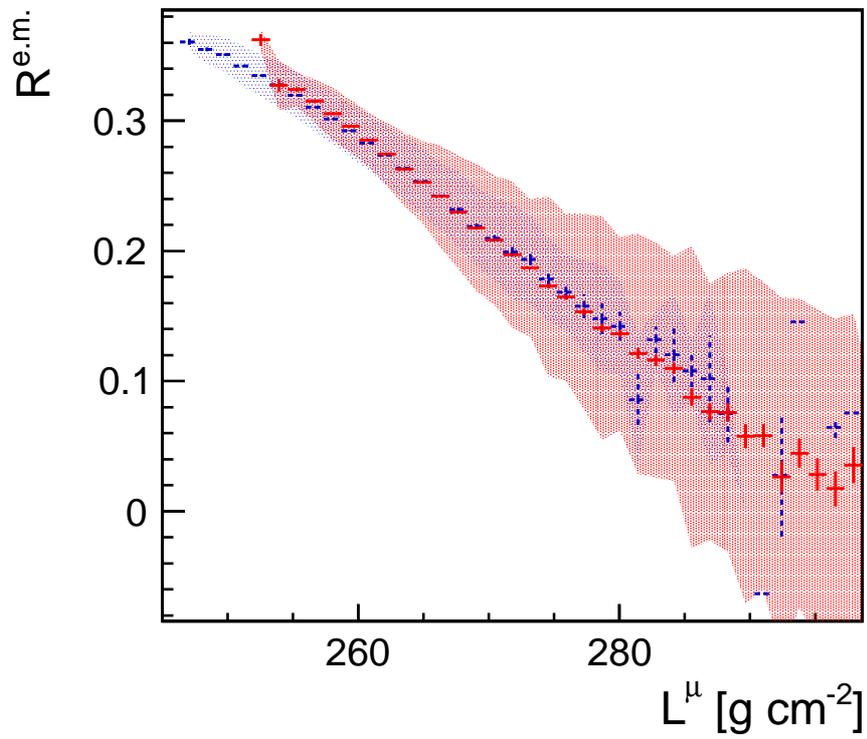}
   \caption{Correlation between the electromagnetic profile shape (characterized by $R^{e.m.}$) and the muon production profile shape (represented by $L^{\mu}$). The results are shown for proton (red/full) and iron (blue/dashed) induced showers,  at E = $10^{19}$ eV, and with QGSJet-II. The shaded area shows the one region with one sigma.}
   \label{fig:shapecorr}
 \end{center}
\end{figure}

The relation between the shapes of the electromagnetic and the muon production profile of each individual event can be seen in fig. \ref{fig:shapecorr}. Here, $(L^{e.m.}, R^{\mu})$ are fixed to its corresponding average values, and most of the information is kept in a single, most sensitive variable. The correlation is rather strong in the most populated region of $(R^{e.m.}, L^{\mu})$. Moreover, it is almost independent of the primary mass composition, making it very useful for hybrid analysis. Indeed, whenever one of the profiles is measured accurately a prediction of the other profile shape can be established.




\section{Dependence on the energy cutoff}
\label{sec:Ecut}

The previous studies were done with CONEX generated profiles, including a muon energy threshold\footnote{minimum value permitted in CONEX simulations.} of 1 GeV.
The final threshold for observation depends not only on the detector used, but also on the zenith angle of the shower, and within the same shower will be different for different depths. 
It is worth verifying how do the previous conclusions depend on the threshold used.

\begin{figure}[htbp]
 \begin{center}
  \subfigure[]{
   \includegraphics[width=0.45\textwidth]{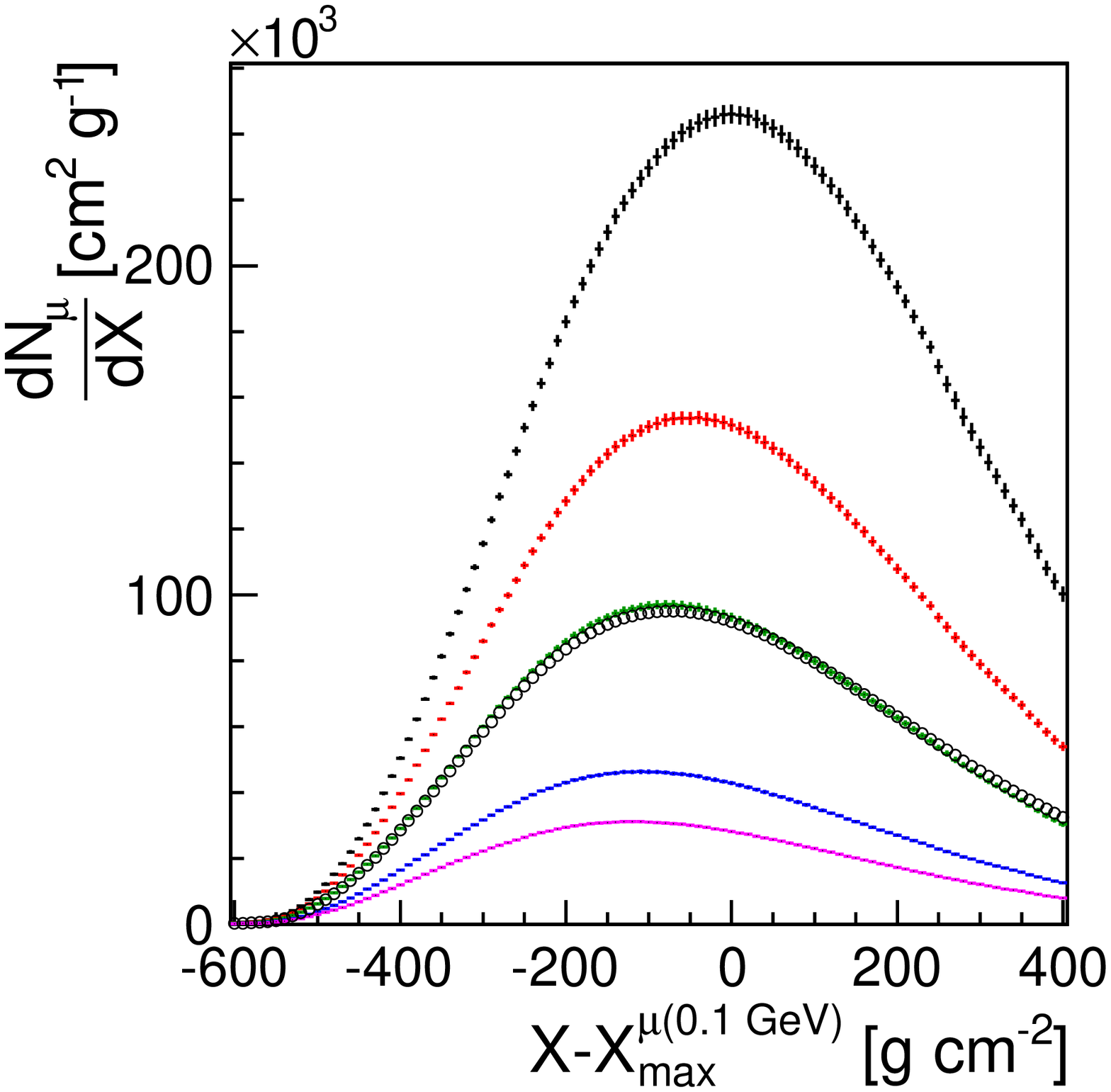}
    \label{fig:Ecutoff_norm}
  }
   \hspace{0.7cm}
  \subfigure[]{
   \includegraphics[width=0.45\textwidth]{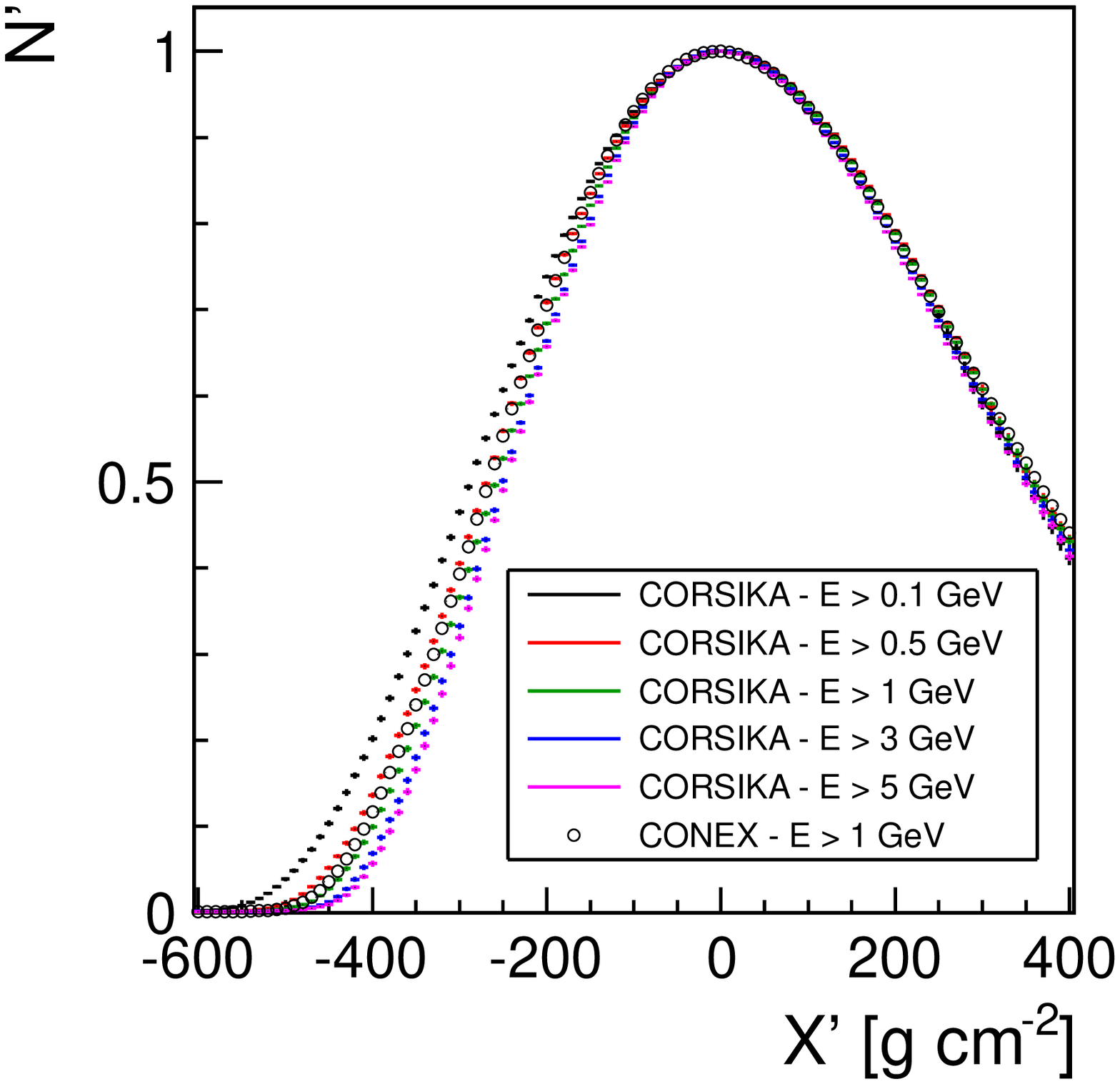}
    \label{fig:Ecutoff_usp}
  }
   \caption{Average muon production profile as a function of the muon energy threshold, for proton induced showers at E = $10^{19}$ eV, with QGSJet-II. The CORSIKA showers are the average of $100$ events while CONEX has $50000$ events. In (a) the profiles are displaced to the maximum depth of the profile with $E^{\mu}_{th} > 0.1$ GeV. In (b) average profiles are shown in $(X',N')$ coordinates.}
   \label{fig:profth}
 \end{center}
\end{figure}

Fig. \ref{fig:profth} is done for proton showers generated with CORSIKA \cite{CORSIKA} (version 6.980), which allows the setting of lower energy thresholds (but is much slower, so there is less statistics). This version was modified in order to obtain the muons at production.
We first checked that the results obtained with CORSIKA and CONEX are compatible at 1 GeV. From Fig. \ref{fig:Ecutoff_norm} is clear that $N^{\mu}_{max}$ changes drastically with the cut considered, and also $X^{\mu}_{max}$ has some variation, since low energy muons
can be produced until much later.  
However all the profiles are still described by a Universal Shower Profile, i.e. using eq. \ref{USPVform}, which can still be described by $L^{\mu}$ and $R^{\mu}$. 
From Figure \ref{fig:Ecutoff_usp} we can infer that $L^{\mu}$ gets smaller as the energy increases, and the $R^{\mu}$ gets slightly larger. 

To extract the full information about the muon production profile, there will be need to invert the propagation effects, taking into account the muon energy spectra and transverse momentum \cite{TransportModel}. It will also be useful to have a detector which can select muons with a minimum energy threshold. The systematic uncertainties coming from the translation between the modeled distribution and the detected one will thus be minimized. These systematic uncertainties can be estimated directly with the data by comparing different detection conditions, selected by zenith angle bins or by individual detectors seeing the same event. In any case, the ideas of universality and the new composition variables are still valid, even if the method has to be calibrated for different detection conditions using data.

\section{Total Number of Muons and Muons from the First Interaction Point}
\label{sec:NbMuons}

The integral of a Gaisser-Hillas profile is readily obtained from $\int N^{\mu} dX = \sqrt{2 \pi} \cdot N_{max} \cdot L^{\mu} \cdot f(R^{\mu})$, in which $f(R^{\mu})$ is a small correction of around $1.02$ for $R^{\mu}\sim$ 0.5. 
This gives the energy in electromagnetic profiles, and the total number of produced muons (above a given energy threshold) for the muon production profiles. 
So, by using the above formalism, the total number of muons can be obtained with a small uncertainty, just by detecting the region of shower maximum, in an event-by-event way. Moreover, as seen before, at first order, $L^{\mu}$ is constant, and consequently $N^{\mu}_{max}$ can be used itself as a measurement of the total number of muons produced during the shower development.

This is reasonably different from the usual muon counting methods, which deal with the muons observed at ground, more dependent on detection conditions and harder to relate to the underlying physics. 
Of course, for this the muon transport model has to take into account the produced muon spectrum, keep track of the muon decays through the atmosphere and detection efficiencies have to be considered.

On the other hand, when working in X', the shape variables $L$ and $R$ become a measurement of $\Delta X$; together with $X_{max}$ they can be used to calculate the point of first interaction, $X_1=X_{max}-\Delta X$. For the muon production profile, $R^{\mu}$ can be fixed, and $L^{\mu}$ can be measured close to the shower maximum, so that at least a part of the variation on $\Delta X$ can be accounted for most of the measured events. In addition, the MPD distribution starts much earlier than the energy deposit profile with muons being produced directly from the first few interactions and with a much steeper rise. 


The fraction of produced muons that will decay before reaching the ground level increases with the zenith angle $\theta$. Hence, the energy threshold for the muon to reach the ground depends also on this quantity, $\theta$. Studies of the distribution of the first muons as a function of $\theta$ can thus give some insight about the parent pion energy distributions in the high energy interactions.


\begin{figure}[htbp]
 \begin{center}
   \includegraphics[width=0.8\textwidth]{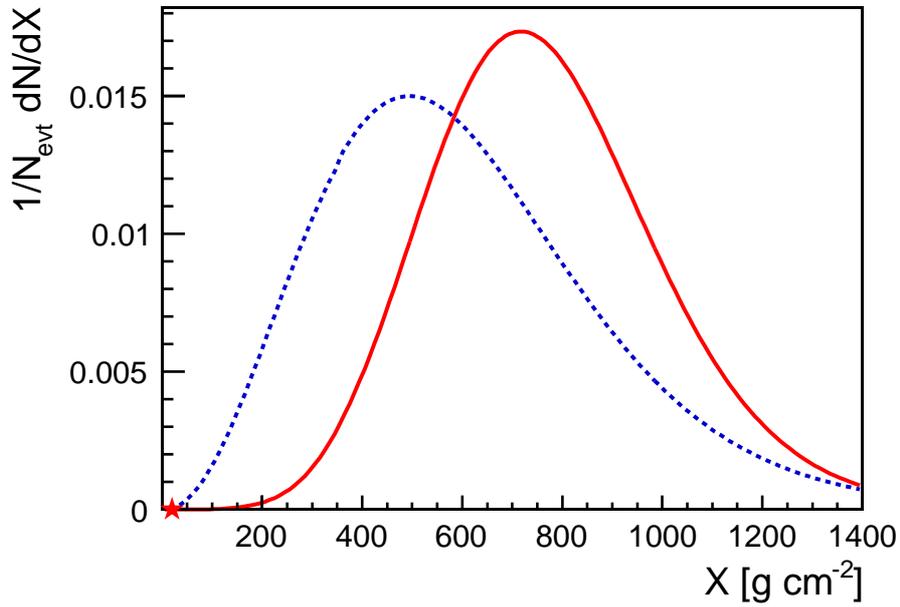}
   \caption{Electromagnetic (red/full) and muon production (blue/dashed) longitudinal profiles of a single \emph{typical} proton induced EAS at E = $10^{19}$ eV. The shower was produced using QGSJet-II as high energy hadronic interaction model. Both profiles were normalized to the area. The first interaction point is shown as a red star.}
   \label{fig:prof2}
 \end{center}
\end{figure}



\section{Conclusions and Prospects}
\label{sec:Conclusion}

The muon production longitudinal profile of air showers is a true characteristic of the shower (independent of detection conditions), described by the same parametrization as the electromagnetic profile. It gives new primary mass composition variables which are fairly independent of the high energy hadronic interaction  model: the $X^{\mu}_{max}$ and the shape variable $L^{\mu}$. These variables can be combined to obtain the point of first interaction. The normalization of the profile gives also access to the total number of produced muons, which is known to be an important variable, both for primary composition and high energy hadronic interaction model studies. 


Thus the muon production profile enclosures important information about the primary composition provided that the experimental systematics are under control, as well as muon energy spectrum, transverse momentum and other elements that play a role on the muon transport.

Joining all the information with the electromagnetic profile will give rise to extra variables, mostly sensitive to the shower development characteristics, and to a more precise understanding in terms of energy distribution along the shower development. With the higher number of available observables, using profiles which are independent of the detection conditions and more directly related to the hadronic cascade, cosmic rays become increasingly useful for the study of particle physics at the highest energies.

\section*{Acknowledgments}
We would like to thank J. Alvarez-Mu\~niz, R. Engel and the Auger LIP group for careful reading the manuscript. This work is partially funded by Funda\c{c}\~{a}o para a Ci\^{e}ncia e Tecnologia (CERN/FP11633/2010 and SFRH/BPD/73270/2010), and fundings of MCTES through POPH-QREN-Tipologia 4.2, Portugal, and European Social Fund.


\bibliographystyle{elsarticle-num}
\bibliography{Bib-hX}

\end{document}